\documentstyle[11pt,epsfig]{article}

\def\laq{~\raise 0.4ex\hbox{$<$}\kern -0.8em\lower 0.62
ex\hbox{$\sim$}~}
\def\gaq{~\raise 0.4ex\hbox{$>$}\kern -0.7em\lower 0.62
ex\hbox{$\sim$}~}

\def\vp{\varphi}
\def \vpb {{\overline {\vp}}}

\def \rb {\overline \rho}
\def \pb {\overline p}

\begin{document}

\begin{titlepage}

\begin{flushright}
CERN-PH-TH/2004-198
\end{flushright}

\vspace*{1.8 cm}

\begin{center}

\huge
{Rotational inhomogeneities from pre-big bang?}

\vspace{1cm}

\large{Massimo Giovannini}

\normalsize
\vspace{.2in}

{\sl Department of Physics, Theory Division, CERN, 1211 Geneva 23, Switzerland}

\vspace*{1.5cm}
\begin{abstract}
The evolution of the rotational inhomogeneities is investigated in the 
specific framework of four-dimensional pre-big bang models.
While minimal (dilaton-driven) scenarios do not lead to rotational 
fluctuations, in the case of non-minimal (string-driven) 
models, fluid sources are present in the pre-big bang phase. The 
rotational modes of the geometry, coupled to the divergenceless 
part of the velocity field, 
can then be amplified depending upon the value of the barotropic index of the 
perfect fluids. In the light of a possible production of rotational 
inhomogeneities, solutions describing the coupled evolution of the 
dilaton field and of the fluid sources are scrutinized in both the string 
and Einstein frames.
In semi-realistic scenarios, where the curvature divergences are regularized 
by means of a non-local dilaton potential, the rotational inhomogeneities are 
amplified during the pre-big bang phase but they decay later on. 
Similar analyses can  also be performed  when a contraction 
occurs directly in the string frame metric.
\end{abstract}
\end{center}

\end{titlepage}
\newpage

\renewcommand{\theequation}{1.\arabic{equation}}
\setcounter{equation}{0}
\section{Introduction}
If a four-dimensional Friedmann--Robertson--Walker (FRW) Universe 
is expanding, the rotational fluctuations of the geometry are not amplified.
This statement is well known since the early analyses of  
Lifshitz and Khalatnikov (see section 9 of Ref. \cite{LK}).

The results of Ref. \cite{LK} assume four important premises:
\begin{itemize}
\item{} the Universe is {\em four-dimensional} and it is 
described by a FRW line element;
\item{} the matter sources driving the evolution of the geometry are 
{\em perfect relativistic fluids} with barotropic equation of state;
\item{} the Universe is always {\em expanding};
\item{} the connection between the evolution of the matter sources 
and the evolution of the geometry is provided by {\em general relativity}.
\end{itemize}

If these four premises are rigorously verified the conclusions
of Ref. \cite{LK} follow by studying the evolution of the rotational
inhomogeneities of the metric in a FRW background. The rotational 
inhomogeneities of the metric are parametrized in terms of two divergenceless
vectors in three spatial dimensions, leading, overall, to four independent 
degrees of freedom (see section 2 below
 for a more formal discussion).
The rotational fluctuations of the geometry are also coupled, through
Einstein equations, to the divergenceless part of the velocity 
field supported by the presence of a perfect barotropic fluid.

If some of the mentioned premises are not verified, there is 
the logical possibility that the rotational inhomogeneities are amplified.
For instance Grishchuk \cite{GR} analysed the situation where the Universe 
is not filled by a perfect barotropic fluid. Indeed, in the model 
of Ref. \cite{GR}, the early stages of the life of the Universe 
are described by a medium characterized by a non-vanishing torque force. 
The resulting rotational  modes are, in this case, copiously produced.

Looking at the hypotheses listed above, there could also be, in principle,  
other ways of obtaining rotational modes of the geometry and of the 
velocity field in reasonable cosmological scenarios. For instance, it could 
happen that the geometry is  not purely four-dimensional or that 
the dynamical equations connecting the evolution of the 
matter sources to that of the geometry do not follow 
from general relativity but from some other non-Einsteinian 
theory of gravity.

Indeed, if the geometry has more than four dimensions, the rotational 
degrees of freedom are more numerous than in the four-dimensional case.
Furthermore, depending on the specific theory, the evolution equations 
of the rotational degrees of freedom are qualitatively different 
with respect to the four-dimensional case.
This possible perspective was recently invoked in \cite{G1}, where 
the evolution of the vector modes of a five-dimensional theory 
have been analysed in the context of a regular model of dynamical 
dimensional reduction.

Yet another way of amplifying dynamically rotational inhomogeneities
would be to assume that the Universe was not always expanding 
but that there have been  long periods of contraction.
Different (unconventional) paradigms of the early Universe 
may realize, in one way or  another, this requirement.
For instance, various incarnations of string cosmological scenarios, such as 
the pre-big bang \cite{V1,V2,G2} and the ekpyrotic \cite{T1,T2,T3}
paradigms, assume that in the past history of the Universe a rather 
long contracting phase took place. To be more specific, the 
pre-big bang dynamics is described by a phase of accelerated expansion in the 
string\footnote{The string and Einstein frames are 
two different parametrizations of the action 
of the theory and are connected by simple 
(local) field redefinitions involving the metric and the dilaton field.
In particular, in the string frame action (see section 2) the dilaton field 
is directly coupled to the Ricci scalar, while in the Einstein frame action 
the dilaton is not directly coupled to the Einstein--Hilbert 
term (see section 3). Notice that, as far as the description is concerned, both 
the Einstein and the string frames can be employed for actual calculations 
and the results will be the same. However, at a practical level, 
 there are cases  where one frame 
is more convenient than the other. Interesting arguments in 
favour of the string frame were put forward since 
the centre of mass of fundamental test strings follows 
geodesics in the string and not in the Einstein frame \cite{H1}.} 
frame but this evolution becomes, indeed, an accelerated 
contraction in the more conventional Einstein frame \cite{V3,V4}.
On the contrary, in some explicit realization of the ekpyrotic scenarios, 
a slow contraction takes place already in the string frame 
description.

Concerning the issue raised in the 
previous paragraph,  a relevant technical remark is in order.
The possibility of defining different frames is inherent to the 
non-Einsteinian nature of the low-energy string effective action. 
It should be borne  in mind that the first three of the four assumptions 
mentioned at the beginning of this investigation are not independent 
in the sense that the word {\em contraction} has a 
definite meaning only if the action of the underlying theory 
is specified. Thus, provided the theory is not of Einstein-Hilbert 
type, there is no surprise if the rotational modes 
are amplified in an expanding Universe.

There is an important physical distinction to be made when 
discussing the amplification of fluctuations in a cosmological context. 
The amplification can be either {\em adiabatic} or {\em 
super-adiabatic}. The adiabatic amplification implies the growth of the 
amplitude. The super-adiabatic amplification implies that the evolution 
equations exibit an instability for a given range of Fourier modes. 
If the vector modes are the ones of a four-dimensional Universe, the 
equations of motion, as it will be shown, only allow the possibility 
of adiabatic amplification (unless specific couplings to som medium  with 
non-vanishing torque force are envisaged as in Ref. \cite{GR}).
If, however, we go to higher dimensions (for instance in $5$ space-time 
dimensions), there, one more physical vector mode appears;
its evolution  exhibits clearly the phenomenon of 
super-adiabatic amplification \cite{G2}.

Recently, Battefeld and Brandenberger \cite{BB} argued that vector modes 
of the geometry can be produced in contracting models. 
To be more specific, the argument of \cite{BB} assumes that the matter 
content of the Universe is {\em only} given by a perfect relativistic fluid and 
that the theory is of Einstein-Hilbert type.
If the matter content of the Universe 
would consist only of a scalar field driving a contraction, the effective 
velocity field will be automatically irrotational. This is exactly the 
situation of minimal pre-big bang models. 

The authors of Ref. \cite{BB} did not 
consider any specific regular solution connecting, in their framework, 
the contracting to the expanding phase. 
This remark has already been introduced in \cite{G1} where it was, almost incidentally, 
 argued that a complete study of string-driven 
pre-big bang models should be performed in order to support the idea of 
amplification of the rotational modes of the geometry. In this spirit, one 
of the sections of \cite{G1} dealt with a simplified four-dimensional 
model whose analysis showed that vector modes of the geometry 
may well be amplified but decay later on.

The motivation of the present paper is to present a plausible
 systematic analysis  
of the possible production of rotational inhomogeneities 
in four-dimensional pre-big bang models. 
String cosmological models incorporating the evolution of fluid sources 
are often qualified string-driven since, in these cases, inflation can 
be obtained also by studying the evolution of a gas of fundamental strings 
\cite{V5}.

Before plunging into the analysis it is appropriate 
to mention some investigations dicussing rotational fluctuations 
in a perspective different from the one employed in the present paper.
In Ref. \cite{BD}, G\"odel solutions have been investigated in the low-energy
string effective action including string tension corrections. In these 
solutions the magnitude of the vorticity vector is proportional to the inverse 
of the string tension. 
In \cite{chm1,chm2,chm3} where the 
evolution of the vorticity was studied in Friedmann-Robertson-Walker models.
The approach followed by these authors is to perturb around a spherically 
symmetric profile of the geometry. The approach reported in the present investigation
is connected with the Bardeen formalism and it is then different from the one of 
 Ref. \cite{chm1,chm2,chm3} . The results of the two approaches 
 seem to be in agreement since in Ref. \cite{ chm1,chm2,chm3} 
 is also found that the rotation parameter
depends upon an inverse power of the scale-factor . 
Finally, in \cite{ocs} the possible r\^ole of rotation was investigated in the 
context of Bianchi type-IX models. 

It is also appropriate to mention some early (very interesting) 
references dealing with stiff fluids in the early Universe. In \cite{B1a} 
(and, partially also in \cite{B2a})  one of the unusual 
features of fluids stiffer than radiation was pointed out: the 
rotational velocity increases as the Universe expands. A reprise 
of this theme appears to be Ref. \cite{EB} where similar considerations 
are discussed in the context of the so-called holographic cosmology.

The present study is organized as follows. In section 2  
various (singular) solutions of the low-energy  string effective 
action will be investigated in the presence of perfect 
fluids with barotropic equation of state.  The evolution of the 
rotational inhomogeneities will then be obtained. In section 3 
the Einstein frame picture will be presented in the light 
of the evolution of the rotational modes.  
Then, in section 4, non-singular models of pre-big bang will be 
specifically discussed and the corresponding evolution 
of the rotational inhomogeneities derived.
Finally, section 5 contains some concluding (critical) remarks.

\renewcommand{\theequation}{2.\arabic{equation}}
\setcounter{equation}{0}
\section{Pre-big bang and ekpyrotic initial conditions}

Consider, to begin with, the low-energy string effective action in the 
presence of fluid sources in four space-time dimensions:
\begin{equation}
S = - \frac{1}{\lambda_{\rm s}^2} \int d^{4} x \sqrt{- G}  e^{- \varphi} [ R 
+ (\partial \varphi)^2 + W(\varphi)] + S_{\rm m}.
\label{action1} 
\end{equation}
In Eq. (\ref{action1}) $\lambda_{\rm s}$ is the string length scale 
and $W(\varphi)$ is a (local) potential depending on the four-dimansional 
dilaton field 
$\varphi$; the term $S_{\rm m}$ represents the contribution 
to the total action coming from perfect barotropic fluids. The compact 
notation 
\begin{equation}
(\partial \varphi)^2 = G^{\alpha\beta} \partial_{\alpha}\varphi \partial_{\beta} \varphi,\,\,\,\,\,\,\, \nabla^2 \varphi = G^{\alpha\beta} \nabla_{\alpha} 
\nabla_{\beta} \varphi = 
G^{\alpha\beta}( \partial_{\alpha}\partial_{\beta} \varphi - 
\Gamma_{\alpha\beta}^{\sigma} \partial_{\sigma} \varphi),
\end{equation}
will be also employed throughout this investigation. The possible coupling 
of the dilaton field to the matter sources as well as the possible 
presence of an antisymmetric tensor field have been neglected. 
As already stressed in the introduction, 
the action (\ref{action1}) has been written in what 
is customarily defined as the string frame metric.

The functional derivation of (\ref{action1}) 
 with respect to $G_{\mu\nu}$ and $\varphi$  leads to the 
well-known equations:
\begin{eqnarray}
&& {\cal G}_{\mu\nu} + \nabla_{\mu}\nabla_{\nu} \varphi + 
\frac{1}{2} G_{\mu\nu}[ (\partial\varphi)^2 - 2 \nabla^2 \varphi - W(\varphi)]
= \lambda_{s}^2 e^{\varphi} T_{\mu\nu},
\label{eq1}\\
&& R + 2 \nabla^2 \varphi - (\partial\varphi)^2 + W - 
\frac{\partial W}{\partial \varphi} =0,
\label{eq2}\\
&& \nabla_{\mu} T^{\mu\nu} =0,
\label{eq3}
\end{eqnarray}
where $T_{\mu\nu}$ is the energy-momentum tensor of the fluid sources and 
${\cal G}_{\mu\nu} = R_{\mu\nu} - 1/2 G_{\mu\nu} R$ the usual Einstein tensor.

Combining Eqs. (\ref{eq1}) and (\ref{eq2}) the following useful equation
\begin{equation}
R_{\mu\nu} + \nabla_{\mu} \nabla_{\nu} \varphi 
- \frac{1}{2} \frac{\partial W}{\partial \varphi} G_{\mu\nu} =
\lambda_{\rm s}^2  e^{\varphi} T_{\mu\nu}
\label{eq4}
\end{equation}
can be obtained.
In a spatially flat FRW geometry,
\begin{equation}
ds^2 = G_{\mu\nu} dx^{\mu} dx^{\nu} = dt^2 - a^2(t) d\vec{x}^2,
\label{metric1}
\end{equation}
with homogeneous dilaton field $\varphi= \varphi(t)$, the 
explicit form of the background equations follows from the components 
of Eq. (\ref{eq1}) (or (\ref{eq4})) and  from Eqs. (\ref{eq2}) and (\ref{eq3}):
\begin{eqnarray}
&&\dot{\vp}^2 + 6 H^2 - 6 H \dot{\vp} - W = 2 \lambda_{\rm s}^2 \rho e^{\vp},
\label{b1}\\
&& \dot{H} = H \dot{\vp} - 3 H^2 - \frac{\partial W}{\partial \varphi} + 
\lambda_{\rm s}^2 p e^{\vp}, 
\label{b2}\\
&& 2 \ddot{\vp} -6 \dot{H} - 12 H^2 - {\dot{\vp}}^2 + 6 H \dot{\vp} + W 
- \frac{\partial W}{\partial\varphi} =0,
\label{b3}\\
&& \dot{\rho} + 3 H ( \rho + p)=0.
\label{b4}
\end{eqnarray}
In Eqs. (\ref{b1})--(\ref{b4}) the overdot denotes the derivation with 
respect to the cosmic time coordinate $t$, and $\rho$ and $p$ are 
the energy and pressure density of the perfect fluid whose 
energy-momentum tensor has the standard form
\begin{equation}
T_{\mu\nu} = ( p + \rho) u_{\mu} u_{\nu} -  p G_{\mu\nu} .
\end{equation}
Since we shall assume that the fluid is barotropic, i.e. 
$ p = \gamma \rho$, Eq. (\ref{b4}) gives, after direct integration
\begin{equation}
\rho= \rho_{0} a^{- 3 ( \gamma + 1)}.
\label{b5}
\end{equation}

In pre-big bang models the initial conditions are customarily  set for $ t <0$. 
Indeed we shall be looking for solutions of the  system
of Eqs. (\ref{b1})--(\ref{b4}), which  can then 
be parametrized as
\begin{eqnarray}
&& a(t) = (- \tau)^{\alpha},
\label{p1}\\
&& \vp(t) = \vp_0 - \beta \ln{(-\tau)},
\label{p2}\\
&& \rho = \rho_{0} ( - \tau)^{- 3 \alpha(\gamma + 1)},
\label{p3}
\end{eqnarray}
where $\tau = t/t_0$ and where Eq. (\ref{p3}) follows by inserting 
Eq. (\ref{p1}) into Eq. (\ref{b5}).

Consider now the case $W(\vp) =0$.
Inserting Eqs. (\ref{p1})--(\ref{p3}) into Eqs. (\ref{b1})--(\ref{b3}) 
the following set of relations between the parameters must be satisfied:
\begin{eqnarray}
&& 3 \alpha (\gamma + 1) + \beta = 2,
\label{first}\\
&& \beta ( 2 - \beta) + 6\alpha( 1 - 2 \alpha -\beta)  =0,
\label{second}
\end{eqnarray}
and 
\begin{equation}
\rho_{0} =e^{- \vp_{0}}
 \frac{\beta^2 + 6 \alpha^2 + 6 \alpha\beta }{2 \lambda_{\rm s}^2}  
= e^{-\vp_{0}} \frac{\beta(\alpha-1) + 3\alpha^2}{\gamma \lambda_{\rm s}^2}.
\label{third}
\end{equation}
The second equality in Eq. (\ref{third}) is only required 
in the case $\gamma \neq 0$.
In fact, in the case $\gamma =0$ Eqs. (\ref{first}) and (\ref{second}) 
imply $\alpha =0$ and $\beta =2$. In this solution 
the scale factor and the energy density are both constant while 
the dilaton evolves in time. 

If $\beta = 0$, Eq. (\ref{p2}) implies that the dilaton is constant. 
According to Eq. (\ref{first}): 
\begin{equation}
\alpha = \frac{2}{ 3 (\gamma +1)}.
\label{alpha1}
\end{equation}
This is the standard FRW solution obtainable in a general relativistic context 
when fluid sources are present. 

Finally, if $\gamma \neq 0$,  the full solution 
will be 
\begin{eqnarray}
&& a(t) = (- \tau)^{\frac{2\gamma}{1 + 3 \gamma^2}},
\nonumber\\
&& \vp(t) = \vp_{0} + \frac{6 \gamma -2}{1 + 3 \gamma^2} \ln{(-\tau)},
\nonumber\\
&& \rho(t) = \rho_{0} ( - \tau)^{- \frac{6 \gamma(\gamma + 1)}{1 + 3 \gamma^2} }.
\label{alpha2}
\end{eqnarray}
The solution (\ref{alpha2}) describes 
an accelerated expansion ( $\dot{a} >0$ and $\ddot{a} >0$) {\em provided}
$ \gamma <0$.  In this case the solution (\ref{alpha2}) describes 
the initial, i.e. $t \to - \infty$, asymptotic state typical of the 
string-driven pre-big bang scenario \cite{V4}. 
Conversely, if $\gamma >0$ the solution contracts ($\dot{a} <0$ and 
$\ddot{a} <0$) in the string frame metric. This situation is 
reminiscent of the initial asymptotic state of ekpyrotic scenarios. 

If the dilaton is constant the solution, expressed by Eqs. (\ref{p1}) and 
 (\ref{alpha1}), 
contracts for all $ -1 < \gamma \leq 1$  and for $ t \to -\infty$. 
Incidentally, notice that a particular subset of these solutions 
(i.e. where $\gamma =1$) is the one used in \cite{BB}
to argue that rotational  inhomogeneities are 
amplified in pre-big bang models.  

Of course, the evolution of the solutions described so far 
can only be applied for $ t < 0$, i.e. during the pre-big bang 
phase. What happens next is determined by the way
the singularity problem is addressed which we will do in section 4. 
For the moment the attention will be concentrated on the 
evolution of the vector modes during the pre-big bang phase.

Defining, as usual, a conformal time coordinate $\eta$ (related to the 
cosmic time as $a(\eta) d\eta = d t$), the fluctuations of the metric 
(\ref{metric1}) are 
\begin{eqnarray}
&& \delta G_{0i} = - a^2(\eta) Q_{i},
\nonumber\\
&& \delta G_{i j} = a^2(\eta) ( \partial_{i} W_{j} + \partial_{j} W_{i}),
\end{eqnarray}
with $ \partial_{i} Q^{i} = \partial_{i} W^{i} =0$. For infinitesimal 
coordinate transformations 
\begin{equation}
x^{i} \to \tilde {x}^{i} = x^{i} + \zeta^{i},
\label{VT}
\end{equation}
preserving the vector nature of the fluctuation (i.e. 
$ \partial_{i} \zeta^{i}=0 $ ), the two variables $Q_{i}$ and $W_{i}$ 
shall change according to the following transformation rules 
\begin{eqnarray}
&& \tilde{Q}_{i} = Q_{i} - \zeta_{i}' ,
\label{QI}\\
&& \tilde{W}_{i} = W_{i} + \zeta_{i},
\label{WI}
\end{eqnarray}
where the prime denotes the derivation with respect to the 
conformal time coordinate $\eta$.

From Eqs. (\ref{QI}) and (\ref{WI}) it can be argued that the quantity 
$ \tilde{W}_{i}' + \tilde{Q}_{i}$ is invariant under infinitesimal 
coordinate transformations (\ref{VT}) preserving the vector nature of the 
fluctuation.  Such a quantity is the rotational generalization 
of the so-called Bardeen potentials \cite{BARD},
 which are customarily defined in the case 
of scalar fluctuations of the geometry.

With these specifications  it is  useful to perform calculations in a  gauge. 
There are two possible natural gauge choices.  A first choice could be to set 
$\tilde{Q}_{i} =0$. In this case the gauge function $\zeta_{i}$ is determined 
to be 
\begin{equation}
\zeta_{i}(\eta,\vec{x})  = \int^{\eta} Q_{i}(\eta',\vec{x}) d \eta' + C_{i}(\vec{x}).
\end{equation}
Since the gauge function is determined up to an arbitrary (space-dependent) 
constant, this choice of gauge does not completely fix the coordinate system 
and this occurrence is reminiscent of what happens in the synchronous 
coordinate system of scalar fluctuations \cite{BARD}. The gauge choice 
$\tilde{Q}_{i} =0$  was used, for instance, by the authors of Refs. \cite{LK}
and \cite{GR}.

Another equally useful choice is the one for which $ \tilde{W}_{i}=0$. 
From Eq. (\ref{WI}) the gauge function is determined as
 $\zeta_{i} = - W_{i}$ and the gauge freedom, in this case, is completely 
fixed. Moreover, in this gauge, the gauge-invariant ``Bardeen'' 
potential coincides with $Q_{i}$ and  the only non-vanishing 
entry of the perturbed metric is $\delta G_{0i}$. Using the fact that 
$\delta G^{0i} = - Q^{i}/a^2$ the Christoffel connections are, to first-order in the 
amplitude of the rotational fluctuations of the geometry:
\begin{eqnarray}
&& \delta \Gamma_{i 0}^{0} = {\cal H} Q_{i}, \,\,\,\,\,\,\, 
\delta \Gamma_{i j}^{k} = - {\cal H} Q^{k} \delta_{ij},\,\,\,\,\,\,
\delta \Gamma_{0 0}^{i} = {Q^{i}}' + {\cal H} Q^{i}
\nonumber\\
&& \delta\Gamma_{i j}^{0}= - \frac{1}{2} ( 
\partial_{i} Q_{j} + \partial_{j} Q_{i}),\,\,\,\,\,\, \delta \Gamma_{i 0}^{j} 
=\frac{1}{2} ( \partial_{i} Q^{j} - \partial^{j} Q_{i}),
\label{christoffel}
\end{eqnarray}
 where, as usual,  ${\cal H} = (\ln{a})'$. The first-order form of the Ricci tensors 
then becomes:
\begin{eqnarray}
&&\delta R_{0 i} 
= ({\cal H}' + 2 {\cal H}^2 ) Q_{i} - \frac{1}{2} \nabla^2 Q_{i},
\nonumber\\ 
&& \delta R_{ij} = - \frac{1}{2} [ (\partial_{i} Q_{j} + \partial_{j} Q_{i})'
+ 2 {\cal H} ( \partial_{i} Q_{j} + \partial_{j} Q_{i}) ].
\label{ricci}
\end{eqnarray}
Finally the first-order form of the Einstein tensors will be, in our gauge,
\begin{equation}
\delta {\cal G}_{0 i} = - ( 2 {\cal H}' + {\cal H}^2) Q_{i} - 
\frac{1}{2} \nabla^2 Q_{i},\,\,\,\,\,\, \delta {\cal G}_{i j} = \delta R_{i j}.
\label{EF}
\end{equation}

The evolution equations of the vector modes of the geometry can be obtained by 
consistently perturbing, for instance, Eqs. (\ref{eq3}) and (\ref{eq4}).
The first-order fluctuation of Eq. (\ref{eq4}) will be 
\begin{equation}
\delta R_{\mu\nu} - \delta \Gamma_{\mu\nu}^{\sigma} \partial_{\sigma} \varphi 
- \frac{1}{2}  \frac{\partial W}{\partial \varphi} \delta G_{\mu \nu} = 
\lambda_{\rm s}^2 \delta T_{\mu\nu},
\label{deltaeq4}
\end{equation}
while the first-order fluctuation of Eq. (\ref{eq3}) implies 
\begin{equation}
\partial_{\mu} \delta T^{\mu\nu} 
+ \delta T^{\nu\alpha} \overline{\Gamma}^{\mu}_{\mu\alpha} + 
\delta \Gamma_{\mu\alpha}^{\mu} \overline{T}^{\nu\alpha}  + 
\delta T^{\alpha\beta} \Gamma_{\alpha\beta}^{\nu} + 
\overline{T}^{\alpha\beta} \delta \Gamma^{\nu}_{\alpha\beta} =0, 
\label{deltaTmunu}
\end{equation}
where $ \overline{\Gamma}_{\mu\nu}^{\alpha}$ and $ \overline{T}^{\alpha\beta}$ 
denote, respectively, the background values of the connections and of 
the energy-momentum tensor.

Recall now that $ \overline{\Gamma}_{i j}^{0} = {\cal H} \delta_{i j}$, 
$\overline{\Gamma}_{00}^{0} = {\cal H}$ and $\overline{\Gamma}_{0i}^{j} =
{\cal H} \delta_{i}^{j}$; using the results of Eqs. (\ref{christoffel}) 
and (\ref{ricci}), the $(0i)$ and $(ij)$ components of Eq. (\ref{deltaeq4}) 
lead to 
\begin{eqnarray}
&& \nabla^2 Q_{i} = - 2 \lambda_{\rm s}^2 a^2 e^{\vp} 
( p + \rho) {\cal V}_{i},
\label{0i}\\
&& Q_{i}' = ( \vp' - 2 {\cal H}) Q_{i}.
\label{ij}
\end{eqnarray}
To obtain the 
precise form of Eq. (\ref{0i}), Eq. (\ref{b2}) has been 
used so as to eliminate a term proportional to $Q_{i}$. In 
Eq. (\ref{0i}),   ${\cal V}_{i}$ is the divergenceless 
fluctuation of the velocity field,  which arises by perturbing to first-order 
the energy-momentum tensor of the fluid sources, i.e. 
\begin{equation}
\delta T_{0i} = (p + \rho) u_{0} \delta u_{i} - p \delta G_{0 i}.
\label{deltaT0i}
\end{equation}
Recalling now that $u_{0}= a$ (as implied by $G^{\mu\nu} u_{\mu} u_{\nu} =1$)
and defining $\delta u_{i} = a {\cal V}_{i}$, Eq. (\ref{deltaT0i}) 
leads to 
\begin{equation}
\delta T_{0i} = a^2 ( p + \rho ) {\cal V}_{i} + a^2 p Q_{i}, \,\,\,\,\, 
\partial_{i} {\cal V}^{i} =0.
\end{equation}
which is the explicit form used to derive Eq. (\ref{0i}).

Using the same conventions for the various perturbation 
variables, the spatial component of Eq. (\ref{deltaTmunu}) 
implies:
\begin{equation}
{\cal V}_{i}' + (1 - 3 \gamma) {\cal H} {\cal V}_{i} =0.
\label{Vi}
\end{equation}
In summary, the Fourier space version of Eqs. (\ref{0i}), (\ref{ij}) 
and (\ref{Vi}) gives
\begin{eqnarray}
&& k^2 Q_{i} = 2 \lambda_{\rm s}^2 a^2 ( p +\rho) e^{\vp} {\cal V }_{i},
\label{0ik}\\
&& Q_{i}' = ( \vp' - 2 {\cal H}) Q_{i},
\label{ijk}\\
&& {\cal V}_{i}' = ( 3 \gamma -1) {\cal H} {\cal V}_{i},
\label{Vik}
\end{eqnarray}
where, to avoid confusion with the vector indices,  
the subscript $k$ denoting the Fourier mode of each 
variable  has been suppressed.

From Eq. (\ref{ijk}) the rotational fluctuation of the geometry becomes 
\begin{equation}
Q_{i} = c_{i}(k) \frac{e^{\vp}}{a^2},
\label{sol1}
\end{equation}
where $c_{i}(k)$ is a ($k$-dependent) integration constant.
Inserting  Eq. (\ref{sol1}) into Eq. (\ref{0ik}) the velocity field 
becomes 
\begin{equation}
{\cal V}_{i} = \frac{k^2 c_{i}(k)}{ 2 \lambda_{\rm s}^2 a^4 (p + \rho)}.
\label{sol2}
\end{equation}
It is immediate to see that Eq. (\ref{sol2}) is consistent 
with Eq. (\ref{Vik}).

Thanks to the results obtained so far, it is now possible 
to discuss the evolution of the rotational 
fluctuations in the various solutions presented earlier 
in this section.

Consider first the solution for $\gamma=0$. As noted above, in this case 
$ a= a_{0}$ and $\rho= \rho_{0}$ are both 
constants but $ \vp = \vp_0 - 2 \ln{(-\tau)}$.
Hence, Eqs. (\ref{sol1}) and (\ref{sol2}) imply:
\begin{eqnarray}
&& Q_{i} = e^{\vp_{0}} \frac{c_{i} (k) a_0}{\tau^2}, 
\nonumber\\
&& {\cal V}_{i} = \frac{k^2 a_0^2 c_{i}(k)}{2 \lambda_{\rm s}^2 \rho_{0}},
\label{QV1}
\end{eqnarray}
showing that while $Q_{i}$ grows for $\tau \to 0^{-}$, ${\cal V}_{i}$ 
stays constant. Furthermore, also $ ( p + \rho) {\cal V}_{i}$, 
the velocity contribution to the fluctuation of the energy-momentum 
tensor, is also constant.

Consider then the case when the dilaton field is constant (or even absent)
 and the only 
matter sources are represented by a perfect barotropic fluid. 
Hence, recalling Eq. (\ref{alpha1}), the evolution of the rotational modes is 
given by:
\begin{eqnarray}
&& Q_{i} \simeq c_{i}(k) ( -\tau)^{- \frac{4}{3 (\gamma + 1)}},
\nonumber\\
&& {\cal V}_{i} \simeq \frac{ k^2 c_{i}(k)}{ 2 \lambda_{\rm s}^2 \rho_0(1 + \gamma)}
(- \tau)^{\frac{2 ( 3 \gamma -1)}{3 (\gamma + 1)}}.
\label{QV2}
\end{eqnarray}
Hence, in the case described by Eqs. (\ref{alpha1}) and (\ref{QV2}) 
the scale factor contracts and $Q_{i}$ increases for $ -1 <\gamma \leq 1$.
Correspondingly, for $\tau \to 0^{-}$,
 ${\cal V}_{i}$ {\em decreases} if $ 1/3 < \gamma \leq 1$ while 
it {\em increases} for $ -1 < \gamma < 1/3$. Note that the case $ \gamma =1$ 
corresponds to the one analysed specifically by \cite{BB}. Furthermore, since 
in this class of solutions the dilaton is constant, the string and Einstein 
frames coincide. The case $\gamma = 1/3$ is characteristic since $ {\cal V}_{i}$ 
is constant while $Q_{i}$ still increases.

It should be appreciated that in the case described by Eq. (\ref{QV2})  
the contribution of the velocity field to the perturbed energy-momentum tensor 
of the sources becomes
\begin{equation}
(p + \rho) {\cal V}_{i} \simeq (-\tau)^{- \frac{8}{3(\gamma + 1)}},
\label{pertexp}
\end{equation}
i.e. $ ( p + \rho) {\cal V}_{i}$ is sharply increasing for 
$\gamma >0$ and for $\tau \to 0^{-}$. Since $|(p + \rho){\cal V}_{i}|$ 
should always be bounded all along the pre-big bang phase, this 
solution clearly presents another problem. One possible conclusion would 
be to infer a general problem of the scenario. On the other 
hand it is also possible to argue that this type of realization 
of pre-big bang dynamics is definitely too naive because 
of the total absence of a dynamical dilaton field. 

In fact, if  more realistic string-driven solutions are 
considered, the situation changes. Consider, indeed, the  class 
of solutions described in Eq. (\ref{alpha2}). Inserting Eq. (\ref{alpha2}) 
into Eqs. (\ref{sol1}) and (\ref{sol2}) the result will be 
\begin{eqnarray}
&& Q_{i} = c_{i}(k) e^{\varphi_{0}}( - \tau)^{ \frac{2 (\gamma -1)}{1 + 3 \gamma^2}},
\label{Qar}\\
&& {\cal V}_{i} = \frac{k^2 c_{i}(k)}{2 \lambda_{\rm s}^2 \rho_{0}(1 + \gamma)} 
(- \tau)^{\frac{2 \gamma ( 3 \gamma - 1)}{1 + 3 \gamma^2}}.
\label{Var}
\end{eqnarray}
As noticed after Eq. (\ref{alpha2}), in this class of solutions, 
$\gamma <0$ implies a phase of accelerated expansion for $\tau <0$. From Eq. 
(\ref{Qar}) $Q_{i}$ increases for all $ -1 \leq \gamma \leq 1$.
From Eq. (\ref{Var}) it can be also concluded that ${\cal V}_{i}$ {\em decreases}
for $ \gamma <0$ and $ \gamma > 1/3$ while it {\em increases} for 
$ 0 < \gamma < 1/3$. Again, the case $\gamma = 1/3$ is special since 
${\cal V}_{i}$ is constant there.

From Eq. (\ref{Var}) the contribution of the rotational fluctuation 
to the perturbed energy-momentum tensor can be obtained:
\begin{equation}
( p + \rho) {\cal V}_{i} \simeq ( - \tau)^{ - \frac{2 \gamma}{1 + 3 \gamma^2}},
\end{equation}
which decreases for $\gamma <0$ while it increases for $\gamma >0$.

The analysis of the solution (\ref{alpha2}) is interesting since it means that 
the pre-big bang initial conditions do not lead to a divergence in the 
perturbed energy-momentum tensor. On the other hand, in the case of contracting 
initial conditions (i.e. $ \gamma >0$ for $\tau <0$), $|(p + \rho) {\cal V}_{i}|$
is unbounded and possibly divergent in the limit $\tau \to 0^{-}$.

Are the results obtained so far conclusive? In some sense yes since it has 
been shown that, for some specific ranges of the barotropic index, 
different classes of solutions of the low-energy string effective 
action seem to indicate that the rotational 
modes of the geometry and of the fluid sources are led to increase.
However, this conclusion cannot be definite unless the whole 
evolution of the model is carefully specified. The pre-big bang initial conditions must 
be analytically connected with an appropriate post-big bang 
solution where, ideally, the dilaton is stabilized and the Universe 
expands in a decelerated way. 

\renewcommand{\theequation}{3.\arabic{equation}}
\setcounter{equation}{0}
\section{Einstein frame description}
 
The analysis  of the rotational inhomogeneities of the 
geometry and of the fluid sources has been conducted, up to now,
in the so-called string frame, where the action takes the form 
(\ref{action1}). In this section we are going to complement 
the results of section 2 by mentioning the main aspects 
of the Einstein frame analysis. This 
task is simplified by the observation that the dilaton 
only contributes to the evolution of the 
homogeneous background. The dilaton {\em fluctuations}, on the 
contrary, do not affect the evolution equation 
of the rotational inhomogeneities. This aspect 
can be appreciated by considering a well-known analogy 
between a perfect relativistic fluid with stiff equation of 
state (i.e. $ p=\rho$) and a scalar field with negligible 
potential. This analogy also implies that the peculiar 
velocity field must be identified with the spatial 
derivative of the dilaton fluctuations. Notice, 
however, that when rotational inhomogeneities are included 
the analogy with the stiff fluid breaks down. While 
a perfect fluid with stiff equation of state 
can support a divergenceless velocity field, the 
``velocity'' associated with the scalar degree of freedom is always 
proportional to $ \partial_{i} \delta\varphi$, $\delta\varphi $ 
being the dilaton fluctuation. It is then clear that the divergenceless
part of the generalized velocity field is 
always vanishing.

With these necessary specifications, the Einstein frame background solutions 
can be obtained from the string frame solutions by means of 
the following local field redefinitions:
\begin{eqnarray}
&&\tilde{G}_{\mu\nu} = e^{- \varphi} G_{\mu\nu},
\label{gtilde}\\
&& \tilde{T}_{\mu}^{\nu} = e^{2 \varphi} T_{\mu}^{\nu},
\label{Ttilde}
\end{eqnarray}
while the dilaton does not change in the transformation between the 
two frames. The  quantities 
with tilde refer to the Einstein frame while the quantities 
without tilde refer to the string frame.

From Eqs. (\ref{gtilde}) and (\ref{Ttilde}) it easily follows that 
\begin{eqnarray}
&& \tilde{a} = e^{- \varphi/2} a,
\label{scalef}\\
&& \tilde{\rho} = e^{2 \varphi} \rho,
\label{rf}\\
&& \tilde{p} = e^{2 \varphi} p.
\label{pf}
\end{eqnarray}
Since, as discussed above, the dilaton fluctuations do not 
contribute to the evolution equations of the vector modes of the geometry, 
from the evolution equations of the fluctuations 
in the string frame, i.e. Eqs. (\ref{0ik}) and (\ref{ijk}), the Einstein frame 
equations can be obtained by identifying the perturbation variables 
through Eqs. (\ref{gtilde}) and (\ref{Ttilde}).
In formulae we will have
\begin{eqnarray}
&& \delta \tilde{G}_{\mu\nu} = e^{-\varphi} \delta G_{\mu\nu},
\label{pt1}\\
&& \delta \tilde{T}_{\mu}^{\nu} = e^{ 2 \varphi} \delta T_{\mu}^{\nu}.
\label{pt2}
\end{eqnarray}
Recall now that $\delta \tilde{G}_{0i} = - \tilde{a}^2 \tilde{Q}_{i}$ and 
$\delta \tilde{T}_{i}^{0} = ( \tilde{p} + \tilde{\rho}) \tilde{{\cal V}}_{i}$.
Hence, it follows from Eqs. (\ref{pt1}) and (\ref{pt2})  that
\begin{equation}
\tilde{{\cal V}}_{i}  = {\cal V}_{i},\,\,\,\,\,\,\, \tilde{Q}_{i} = Q_{i}.
\label{trfl}
\end{equation}
Inserting the result of Eqs.  (\ref{scalef})--(\ref{pf}) and (\ref{trfl}) 
into Eqs. (\ref{0ik})--(\ref{Vik}) the evolution equations 
of the rotational fluctuations can be easily written, in the Einstein frame, as 
\begin{eqnarray}
&& k^2 \tilde{Q}_{i} = (\tilde{p} + \tilde{\rho}) 
\tilde{a}^2 \tilde{{\cal V}}_{i},
\label{fl1}\\
&& \tilde{Q}_{i}' + 2 \tilde{{\cal H}} \tilde{Q}_{i} =0, 
\label{fl2}\\
&& \tilde{{\cal V}}_{i}' = ( 3 \gamma - 1) \biggl( \tilde{{\cal H}} 
+ \frac{\varphi'}{2} \biggr) \tilde{{\cal V}}_{i},
\label{fl3}
\end{eqnarray}
 where $ \tilde{{\cal H}}= \tilde{a}'/\tilde{a}$; units of
$2\lambda_{\rm s}^2 =1$ have been used in Eqs. 
(\ref{fl1})--(\ref{fl3}). An important 
point concerning the derivation of these equations is that 
the conformal time coordinate $\eta$ does not change in the 
transition from the string to the Einstein frame while the cosmic time 
coordinate does change as 
\begin{equation}
d t_{\rm E} = e^{-\varphi/2} dt_{\rm s}.
\label{cc}
\end{equation} 
Using Eqs. (\ref{cc}) and (\ref{scalef})
 and recalling the definition of the 
conformal time coordinate in the two frames (i.e. $ \tilde{a}(\eta_{\rm E})
d\eta_{\rm E} = d t_{\rm E} $ and $ a(\eta_{\rm s})
d\eta_{\rm s} = d t_{\rm s} $)
it can be easily shown, as anticipated, that $\eta_{\rm E} = \eta_{\rm s}$. 

The presence of the dilaton field in Eq. (\ref{fl3}) 
seems to make the whole system incompatible. This is not the case 
since, by virtue of the transformation rules (\ref{Ttilde})  and 
(\ref{rf}),(\ref{pf}), the dilaton will appear explicitly 
in the covariant conservation equation of the background sources, which 
is, in the Einstein frame:
\begin{equation}
\tilde{\rho} ' + 3 {\cal H} ( \tilde{\rho} + \tilde{p}) + 
\frac{\varphi'}{2} ( 3 \tilde{\rho} - \tilde{p}) =0.
\label{consE}
\end{equation}
Equation (\ref{consE}) can be directly obtained from Eq. (\ref{b4}), 
whose explicit form, in the conformal time parametrization is
\begin{equation}
\rho' + 3 {\cal H} ( \rho + p) =0.
\label{consS}
\end{equation}
Inserting Eqs. (\ref{rf}) and (\ref{pf}) into Eq. (\ref{consS}), 
Eq. (\ref{consE}) is immediately obtained since, from Eq. (\ref{scalef}), 
${\cal H} = \tilde{{\cal H}} + \varphi'/2$.

Thus, from Eq. (\ref{fl1}) and (\ref{fl2}): 
\begin{eqnarray}
&&\tilde{Q}_{i} = \frac{\tilde{c}_{i}(k)}{\tilde{a}^2},
\label{QiE}\\
&& \tilde{{\cal  V}_{i}} = \frac{ k^2 \tilde{c}_{i}(k)}{\tilde{a}^4 
(\tilde{\rho} + \tilde{p})} \equiv 
\frac{k^2 \tilde{c}_{i}(k)}{(\gamma + 1) }\tilde{a}^{3\gamma -1} e^{\vp(3\gamma -1)/2}.
\label{ViE}
\end{eqnarray}
The second equality in Eq. (\ref{ViE}) follows by replacing 
$\tilde{\rho}$ with the result of the direct integration of 
Eq. (\ref{consE}).  Equations  (\ref{QiE}) and (\ref{ViE}) 
 are the ones obtained in \cite{BB} in the rather 
specific case of constant dilaton field.

Finally the Einstein-frame version of the solutions with generic barotropic 
index reported in Eqs. (\ref{alpha2}) are 
\begin{eqnarray}
&&\tilde{a} \simeq \biggl( - \frac{\eta}{\eta_{0}}\biggr)^
{- \frac{\gamma -1}{3\gamma^2 - 2 \gamma + 1}},
\nonumber\\
&& \varphi = \tilde{\varphi} \simeq \varphi_{0} + 
 \frac{2(3 \gamma -1)}{3\gamma^2 - 2 \gamma + 1} \ln{
\biggl(- \frac{\eta}{\eta_{0}}\biggr)},
\nonumber\\
&& \tilde{\rho} \simeq \biggl(- \frac{\eta}{\eta_0}\biggr)^{ \frac{6\gamma( 1- \gamma) - 4}{3\gamma^2 - 2 \gamma + 1}},
\label{alpha2E}
\end{eqnarray}
where the normalization constants have been dropped and only the relevant 
time dependence reported.

Let us now compare the evolution of the vector modes of the geometry 
in the case of the solution given in Eq. (\ref{alpha2}) which is written 
in the cosmic time coordinate, i.e. $\tau = t/t_{0}$. This 
solution can be translated in the conformal time parametrization 
by recalling that $a(\eta) d\eta = dt$. Applying the mentioned differential
 relation we find 
\begin{equation}
(-\tau) \simeq (- \eta)^{\frac{3 \gamma^2 + 1}{1 + 3 \gamma^2 - 2\gamma}}.
\label{costocon}
\end{equation}
From Eqs. (\ref{Qar}) and (\ref{Var})  we can then derive the evolution 
of the vector modes in the {\em string frame} and in the conformal time 
coordinate, i.e.
\begin{eqnarray}
&& {\cal Q}_{i} \simeq 
(-\eta)^{\frac{2 (\gamma -1)}{3 \gamma^2 - 2 \gamma + 1}},
\nonumber\\
&& {\cal V}_{i} \simeq 
(- \eta)^{\frac{2 \gamma( 3 \gamma -1)}{3\gamma^2 - 2\gamma + 1}}.
\label{ssol}
\end{eqnarray}
Equation (\ref{ssol}), we repeat,  gives the solution in the string frame. 

Now, let us compute {\em the same evolution} in the Einstein frame. 
Equations (\ref{alpha2E}) are the Einstein frame version 
of Eqs. (\ref{alpha2}). The evolution of the vector modes in the Einstein 
frame is then obtained by inserting Eqs. (\ref{ssol}) into Eqs. (\ref{QiE}) 
and (\ref{ViE}). Comparing the obtained expressions with 
Eq. (\ref{ssol}), we do obtain that 
\begin{equation}
\tilde{{\cal V}}_{i} = {\cal V}_{i},\,\,\,\,\,\,\,\,\,\,
\tilde{Q}_{i} = Q_{i}.
\end{equation}
This is exactly the result anticipated, on a general 
ground, in Eq. (\ref{trfl}). This means that, even if in the two frames 
the evolution of the scale factors is different, the evolution of the 
fluctuations will be the same in the two frames.  The rationale for this result
is that the transformation of the background solution is compensated by the 
transformation of the evolution equations leading, as expected, to the complete 
equivalence of the two approaches.  Notice also, as a final remark, that 
the complete equivalence is also a consequence of the fact that, unlike 
the cosmic time coordinate, the conformal time coordinate {\em is invariant}
when passing from String to Einstein frame (and viceversa). This property 
has been swiftly re-derived below Eq. (\ref{cc}) and is well known.

\renewcommand{\theequation}{4.\arabic{equation}}
\setcounter{equation}{0}
\section{Decaying rotational inhomogeneities}

In order to assess the phenomenological 
relevance of the present exercise as well as of related ideas \cite{BB}, 
it is important to discuss 
what the fate of the rotational inhomogeneities is.  
It should be understood to what extent the rotational modes 
of the geometry will persist during the post-big bang evolution.
A closely  related issue would be to check if the rotational modes of the 
geometry and of the sources may diverge around the time of the 
transition between pre- and post-big bang. 

In the presence of a dynamical dilaton field, it is possible to 
find completely regular solutions describing the transition from the pre- 
to the post-big bang evolution. These  solutions may well 
include fluid sources and can be obtained when the dilaton potential is 
a  function of $e^{-\vpb}$, defined as 
\begin{equation}
e^{- \vpb(x)}= \frac{1}{\lambda_{\rm s}^3} \int d^4 y \sqrt{- G(y)} 
e^{- \varphi(y)} \sqrt{( \partial\varphi)^2_{y}} \delta( \varphi(x) 
-\varphi(y)).
\label{evbp}
\end{equation}
These solutions were explored through  various steps \cite{MV1,MM1}
(see also \cite{V4}).

In \cite{G3} it
 has been shown that a generally covariant action can be written for 
these models. The evolution of the scalar and tensor modes of the geometry has 
been discussed in detail and it has been concluded that the evolution 
of these inhomogeneities is completely regular all along the different stages 
of the model \cite{G4}. In \cite{G5} it was demonstrated that 
back-reaction effects of massless particles can naturally induce a hot 
phase dominated by radiation with a stabilized dilaton field 
(see also \cite{Ga}  for some earlier attempts to stabilize the dilaton 
in a radiation-dominated post-big bang evolution).

In this section we are going to discuss the evolution of the rotational modes 
in the framework of the same regularization scheme as introduced in \cite{G3,G4}.
Consider indeed the action \cite{G3} 
\begin{equation}
S = - \frac{1}{2 \lambda_{\rm s}^2} d^4 x \sqrt{ - G} e^{- \varphi} [ R 
+ (\partial \varphi)^2 + V ] + S_{\rm m},
\label{action2}
\end{equation}
where $ V \equiv V( e^{- \vpb})$. 

Following the results of Refs. \cite{G2,G3} the relevant equations of motion 
can be written as 
\begin{eqnarray}
&&{\cal G}_{\mu\nu} + \nabla_{\mu}\nabla_{\nu} \varphi + 
\frac{1}{2} G_{\mu\nu} [ (\partial \varphi)^2 - 2 \nabla^2 \varphi - V] 
- \frac{1}{2} e^{-\varphi} \sqrt{(\partial \varphi)^2} \gamma_{\mu\nu} I_{1} 
= e^{\varphi} \lambda_{\rm s}^2 T_{\mu\nu},
\label{nl1}\\
&& \nabla_{\mu} T^{\mu\nu} =0,
\label{nl2}
\end{eqnarray} 
where 
\begin{equation}
I_{1} = \frac{1}{\lambda_{\rm s}^3}\int d^4 y ( \sqrt{- G} V')_{y} 
\delta( \varphi(x) - \varphi(y)),
\label{int1}
\end{equation}
and 
\begin{equation}
\gamma_{\mu\nu} = G_{\mu\nu} - 
\frac{\partial_{\mu} \varphi\partial_{\nu}\varphi}{ (\partial\varphi)^2}.
\label{projector}
\end{equation}
In Eq. (\ref{int1}) $V' = \partial V/\partial(e^{-\vpb})$.
In the homogeneous limit the evolution equations of the geometry are 
found to be \cite{G3,G4}
\begin{eqnarray}
&& {\dot{\vpb}}^2 - 3 H^2 - V =2 \lambda_{\rm s}^2 e^{\vpb}\rb
\label{c1}\\
&& \dot{H} = H \dot{\vpb} + \lambda_{\rm s}^2 e^{\vpb} \pb,
\label{c2}\\
&& 2 \ddot{\vpb} - {\dot{\vpb}}^2 - 3 H^2 + V - \frac{\partial V}{\partial \vpb}
=0,
\label{c3}\\
&& \dot{\rb} + 3 H\pb =0.
\label{c4}
\end{eqnarray}
Notice that, in the homogeneous limit, from Eq. (\ref{evpb}),  
$e^{- \vpb} = e^{-\varphi} a^3$, where we have absorbed into $\vp$ the dimensionless
constant $-\ln{(\int d^{3} y /\lambda_{\rm s}^3)}$ associated with the (finite)
comoving spatial volume. In Eqs. (\ref{c1})--(\ref{c4}) the reduced 
energy and pressure densities 
\begin{equation}
\rb= a^3 \rho,~~~~~~~~~~~\pb= a^3 p
\end{equation}
have been defined.

The evolution equations of the vector modes of the geometry 
can be obtained, following the same notation as in section 2, 
from
\begin{equation}
\delta {\cal G}_{\mu\nu} - \delta \Gamma_{\mu\nu}^{\sigma} \partial_{\sigma} 
\varphi + \frac{1}{2} \delta G_{\mu\nu} [ (\partial\varphi)^2 - 
2 \nabla^2 \varphi - V] - \frac{1}{2} e^{-\varphi} \sqrt{(\partial \varphi)^2} 
\delta \gamma_{\mu\nu} I_{1} = \lambda_{\rm s}^2 \delta T_{\mu\nu}.
\end{equation}
The perturbed form of the covariant conservation equation is the same as the 
one reported in Eq. (\ref{deltaTmunu}).  The fluctuations of $I_{1}$ do 
not contribute to the vector fluctuations. Moreover, in the gauge 
$\tilde{W}_{i}=0$, $\delta\gamma_{0i} = - a^2 Q_{i}$. 
Using now  Eqs. (\ref{christoffel}) and (\ref{EF}), the evolution equations of 
the vector modes can be written, in Fourier space, as:
\begin{eqnarray}
&& Q_{i}' = (\vpb' + {\cal H}) Q_{i},
\label{VN1}\\
&& k^2 {\cal V}_{i} = 2 \lambda_{\rm s}^2 ( \rb + \pb) a^2 e^{\vpb},
\label{VN2}\\
&& {\cal V}_{i}' = ( 3 \gamma -1) {\cal H} {\cal V}_{i}.
\label{VN3}
\end{eqnarray}

Equations (\ref{c2}) and (\ref{c3}) can be written, in the case $V = - V_{0} 
e^{2\vpb}$ as 
\begin{eqnarray}
&& \frac{d }{d t} ( H e^{- \vpb}) = \frac{\gamma}{2} \rb,
\label{red1}\\
&& \frac{d }{d t} ( e^{-\vpb} \dot{\vpb}) = - \frac{\rb}{2}.
\label{red2}
\end{eqnarray}
In order to obtain Eq. (\ref{red2}), Eq. (\ref{c1}) has been used to eliminate 
$3 H^2$ from Eq. (\ref{c3}). Notice also that, in Eqs. (\ref{red1}) 
and (\ref{red2}),  the system of  
$2 \lambda_{\rm s}^2 =1 $ units has been adopted.

Let us now define the new time coordinate 
as 
\begin{equation}
\frac{d x}{d t}  = \rb \frac{L}{2},
\label{red3}
\end{equation}
where $L$ is a constant (dimensionful) parameter.
Equations (\ref{red1}) and (\ref{red2}) can be integrated once 
giving 
\begin{eqnarray}
&& H = \frac{\gamma}{L} x e^{\vpb},
\label{H}\\
&& \dot{\vpb} = - \frac{x}{L} e^{\vpb},
\label{dvpb}
\end{eqnarray}
having set to zero the integration constants. 

Inserting Eqs. (\ref{H}) and (\ref{dvpb}) into Eq. (\ref{c1}) 
we will obtain
\begin{equation}
e^{\vpb} = \frac{ \rb L^2}{(1 - 3 \gamma^2) x^2 + V_{0}L^2}.
\label{evpb}
\end{equation}
Clearly $e^{\vpb} $ is positive-definite and non-singular 
iff
\begin{equation}
c^2 = V_0 L^2 >0,\,\,\,\,\,\,\,\,\, b^2= 1 - 3 \gamma^2 >0.
\label{conv}
\end{equation}
This means that the regular solutions obtained in this way 
will be defined for a reasonable set of barotropic indices, namely 
$ |\gamma | < 1/\sqrt{3}$. Inserting  Eq. (\ref{evpb}) into Eqs. 
(\ref{red1}) and (\ref{red2}), and using the new time coordinate $x$ 
as defined in Eq. (\ref{red3}), the full solution can be written as 
\begin{eqnarray}
&& a(x) = ( b^2 x^2 + c^2)^{\frac{\gamma}{ b^2}},
\label{nsa1}\\ 
&& \vpb(x) = \vpb_{0} - \frac{1}{b^2} \ln{(b^2 x^2 + c^2)},
\label{nsa2}\\
&&\rb = \rb_{0} ( b^2 x^2 + c^2)^{- 3\frac{\gamma^2}{b^2}}.
\label{nsa3}
\end{eqnarray}
Now that the explicit solution has been obtained 
 in terms of $x$, the 
relation between $x$ and $t$ can be found from Eq. (\ref{red3}) 
by direct insertion of Eq. (\ref{nsa3}), with the result
\begin{equation}
\frac{ d x }{ d t} = \frac{L\rb_{0}}{2} 
\frac{1}{( b^2 x^2 + c^2)^{3 \frac{\gamma^2}{b^2}}}.
\label{rel}
\end{equation}
From Eq. (\ref{rel}) it also follows that 
\begin{equation}
|b x| \simeq |b t|^{\frac{b^2}{b^2 + 6 \gamma^2}},
\label{xt}
\end{equation}
for $|x| \to\infty$ and $t \to \infty$.

From Eqs. (\ref{0ik}) and (\ref{ijk}) the evolution of the vector 
modes easily follows, using Eqs. (\ref{nsa1})--(\ref{nsa3}):
\begin{eqnarray}
&& Q_{i}  = c_{i}(k) e^{\vpb_0} ( b^2 x^2 + c^2)^{ \frac{\gamma -1}{b^2}}, 
\label{Qinsa}\\
&& {\cal V}_{i} = \frac{ k^2 c_{i}(k)}{(1 + \gamma)\rb_{0}} 
(b^2 x^2 + c^2 )^{\frac{\gamma( 3 \gamma -1)}{b^2}}.
\label{Vinsa}
\end{eqnarray}
In order to understand the evolution of the vector modes 
it should be clear that since $b^2>0$, then $ |\gamma| < 1/\sqrt{3}$.
In this case the exponent appearing at the right-hand side of Eq. 
(\ref{Qinsa}) is always negative, so that $Q_{i}$ increases for $x <0$ and 
decreases for $x>0$, being regular for $x =0$. 
From Eq. (\ref{Vinsa}) it can be argued that the velocity field 
is decreasing for $ x<0$ if (and only if) $ - 1/\sqrt{3} <\gamma <0$ and 
$1/3 < \gamma < 1/\sqrt{3}$. On the contrary, if $ 0 < \gamma < 1/3$, 
${\cal V}_{i}$  increases for $x<0$.

For $ x >0$, ${\cal V}_{i}$ increases for $ - 1/\sqrt{3} < \gamma <0$ and 
$1/3 <\gamma < 1/\sqrt{3}$, while it decreases for $ 0 <\gamma <1/3$.
The case $\gamma = 1/3$ leads to constant ${\cal V}_{i}$.

From the example discussed so far, it seems that for $ -1/\sqrt{3}
<\gamma < 0$  the rotational modes of the sources decrease for $ x < 0$ but 
increase for $x >0$. This effect is a direct consequence of the evolution of 
$ a(x)$ as reported in Eq. (\ref{nsa1}). Notice, indeed, that from Eq. 
(\ref{nsa1})  and (\ref{xt}) $a(t)$ describes an accelerated expansion for 
$t <0$ and $\gamma <0$, but it is contracting for $t>0$ and the same choice 
of $\gamma$. Analogously, if $\gamma >0$, $a(t)$ will contract during 
the pre-big bang phase and it will expand in the post-big bang phase.

It does not seem  realistic to have a contracting evolution in the post-big 
bang phase, which should represent, in some general sense, the early 
phases of our present Universe below the Planck/string energy scale. 

A more realistic example can be formulated in the case of  $ p=0$ and for 
the following potential
\begin{equation}
V = - V_{0} e^{\vpb} - V_{1} e^{4 \vpb}.
\label{mixed}
\end{equation}
Since $p=0$ (and also, clearly, $\pb=0$), Eq. (\ref{c4}) implies 
$\rb =\rb_0$, i.e. that $\rb$ is constant. Thus, Eqs. (\ref{c1})--(\ref{c3}) 
can be integrated to give
\begin{eqnarray}
&& H = \frac{1}{\sqrt{3}} \frac{1}{\sqrt{t^2 + t_0^2} },
\nonumber\\
&& \dot{\vpb} = \frac{t}{t^2 + t_0^2},
\label{real1}
\end{eqnarray}
with $V_{0} = \rb_{0}$ and $t_0^2 = e^{4\vpb_{0}} V_1$.
In the case given by Eq. (\ref{real1}) the evolution of the rotational 
inhomogeneities is given by 
\begin{eqnarray}
&& {\cal V}_{i} = \frac{k^2 c_{i}(k)}{\rb_{0}} ( t + \sqrt{t^2 + t_0^2})^{-
1/\sqrt{3}},
\nonumber\\
&& Q_{i} = e^{\vpb_{0}} \frac{( t + \sqrt{t^2 + t_0^2})^{1/\sqrt{3}}}{
\sqrt{t^2 + t_{0}^2}}.
\label{real2}
\end{eqnarray}
In this case, again, the rotational modes are regular. The variable 
$Q_{i}$ increases for $t<0$ and decreases for $t>0$, being regular for $t=0$.
The variable ${\cal V}_{i}$ is always decreasing. 

An even more realistic example will now be presented. In this 
case the evolution will interpolate between a pre-big bang phase 
for $t\to -\infty $ and a radiation-dominated phase for $t \to +\infty$.
Consider, indeed, the case in which $ \gamma$  is a function of $x$, 
interpolating between $-1/3$ (for $x\to -\infty$) and $ 1/3$ (for 
$x\to +\infty$). Consider, in particular, the case 
\begin{equation}
\gamma(x) = \frac{1}{3} \frac{x}{\sqrt{x^2 + x_1^2}}.
\label{gammax}
\end{equation}
Equations (\ref{red1}) and (\ref{red2}) can also be integrated, in the case of 
the $\gamma(x)$ given in Eq. (\ref{gammax}), with a 
potential $V = - V_0 e^{2\vpb}$; the solution can be written as
\begin{eqnarray}
&& a(x) =  x + \sqrt{x^2 + x_{1}^2} ,
\nonumber\\
&& \vpb = \vpb_{0} - \frac{3}{2} \ln{(x^2 + x_1^2)},
\nonumber\\
&& \rb = \frac{\rb_{0}}{\sqrt{x^2 + x_1^2}},
\label{solre}
\end{eqnarray}
with 
\begin{equation}
V_{0} L^2 = x_1^2 , \,\,\,\,\,\,\rb_{0} L^2 = \frac{2}{3} e^{\vpb_0}.
\end{equation}
From Eq. (\ref{solre}) 
the evolution of the vector modes can be found to be 
\begin{eqnarray}
&&{\cal V}_{i} = \frac{3 k^2 c_{i}(k)}{\rb_{0}} \frac{x^2 + x_{1}^2 }{( x + 
\sqrt{x^2 + x_1^2})( x + 3 \sqrt{x^2 + x_1^2})},
\label{solre2}\\
&& Q_{i} = c_{i}(k) e^{\vpb_{0}} 
\frac{x + \sqrt{x^2 + x_1^2}}{( x^2 + x_1^2 )^{3/2}}
\label{solre3}
\end{eqnarray}
In order to have a clear physical interpretation of the solution, 
we can notice that, in the asymptotic regions, i.e. $|t| \to \infty$,
\begin{eqnarray}
&& - x \simeq \sqrt{-t},\,\,\,\,\,\, {\rm for }\,\,\,t\to -\infty,
\nonumber\\
&& x \simeq \sqrt{t},\,\,\,\,\,\, {\rm for }\,\,\,t\to \infty.
\end{eqnarray}
Then looking at the specific form of 
$a(x)$ as reported in Eq. (\ref{solre}) it can be appreciated that 
$a(t) \simeq (-t)^{-1/2}$ (for $t \to -\infty$ ) and $a(t) \simeq t^{1/2}$ 
(for $t\to +\infty$).  

According to  Eq. (\ref{solre2}), the velocity field always decreases 
 during the 
pre-big bang epoch and also later on. Notice in fact that 
\begin{eqnarray}
&&{\cal V}_{i} \simeq \frac{c_{i} k^2}{\rb_{0}} x^2,\,\,\,\,\,\, {\rm for}
\,\,\,\, x\to -\infty, 
\nonumber\\
&& {\cal V}_{i} \simeq \frac{c_{i} k^2}{\rb_{0}},\,\,\,\,\,\, {\rm for}
\,\,\,\, x\to \infty.
\end{eqnarray}
If we set initial pre-big bang initial conditions for $x \to -\infty$, ${\cal V}_{i}$ 
will then decrease and then set to a constant value.
From Eq. (\ref{solre3}) the variable $Q_{i}$ increases for $x<0$ but it
decreases, as anticipated,  for $x>0$. 

\renewcommand{\theequation}{5.\arabic{equation}}
\setcounter{equation}{0}
\section{Concluding remarks}

In this paper the evolution of the rotational inhomogeneities 
of the geometry and of the fluid sources has been analysed in the 
framework of four-dimensional pre-big bang models, in both the string 
and Einstein frames. 
The main results can be summarized as follow
\begin{itemize}
\item{} in the case of minimal (dilaton-driven models) the 
rotational inhomogeneities are totally absent;
\item{} if fluid sources are added during the pre-big bang 
phase, the solution of the system shows that the rotational 
inhomogeneities can grow for negative values of the barotropic index, while 
for positive values rotational inhomogeneities decay ;
\item{} the fate of the vector  modes has been analysed in various non-singular
models, where the evolution of the scalar and tensor modes 
of the geometry is well defined and non-singular;
\item{} if the evolution of the geometry and of the dilaton 
is regular, then, as expected, 
 the vector modes are never divergent;
\item{} for realistic examples, where the Universe inflates during the pre-big 
bang epoch and turns into radiation later on, 
 the rotational inhomogeneities increase in the pre-big bang, but swiftly 
decay later on.
\item{} 
in this systematic study, all the known string cosmological solutions with 
fluid sources in four space-time dimensions have been scrutinized. 
\end{itemize}

\end{document}